\documentclass[a4paper,preprintnumbers,floatfix,superscriptaddress,aps,prl,twocolumn,showpacs,notitlepage,longbibliography]{revtex4-1}
\usepackage[utf8]{inputenc}
\usepackage[T1]{fontenc}
\usepackage[sc,osf]{mathpazo}
\usepackage{amsmath, amsthm, amssymb,amsfonts,mathbbol,amstext}
\usepackage{graphicx}
\usepackage{dcolumn}
\usepackage{bm}
\usepackage{bbm}
\usepackage{hyperref}
\usepackage{mathtools}
\usepackage{comment}
\usepackage{color}
\usepackage{multirow}

\usepackage[normalem]{ulem}

\def\1{\mathbf{1}}
\def\0{\mathbf{0}}

\DeclareMathOperator{\Tr}{Tr}

\newcommand{\ket}[1]{| #1 \rangle}

\renewcommand{\rho}{\varrho}

\newcommand{\processnext}[1]{%
  \ifx\listfinish#1\empty\else\listact{#1}\expandafter\processnext\fi}

\usepackage{graphicx, xcolor}

\DeclareGraphicsExtensions{.pdf,.png,.jpg}

\makeindex

\begin{document}
\title{Experimental test of quantum causal influences}
\date{\today}

\author{Iris Agresti}
\affiliation{Dipartimento di Fisica - Sapienza Universit\`{a} di Roma, P.le Aldo Moro 5, I-00185 Roma, Italy}

\author{Davide Poderini}
\affiliation{Dipartimento di Fisica - Sapienza Universit\`{a} di Roma, P.le Aldo Moro 5, I-00185 Roma, Italy}

\author{Beatrice Polacchi}
\affiliation{Dipartimento di Fisica - Sapienza Universit\`{a} di Roma, P.le Aldo Moro 5, I-00185 Roma, Italy}

\author{Nikolai Miklin}
\affiliation{International Centre for Theory of Quantum Technologies (ICTQT), University of Gdansk, 80-308 Gdansk, Poland}
\affiliation{Heinrich Heine University D{\"u}sseldorf, Universit{\"a}tsstra{\ss}e 1, 40225 D{\"u}sseldorf, Germany}

\author{Mariami Gachechiladze}
\affiliation{Institute for Theoretical Physics, University of Cologne, 50937 Cologne, Germany}

\author{Alessia Suprano}
\affiliation{Dipartimento di Fisica - Sapienza Universit\`{a} di Roma, P.le Aldo Moro 5, I-00185 Roma, Italy}

\author{Emanuele Polino}
\affiliation{Dipartimento di Fisica - Sapienza Universit\`{a} di Roma, P.le Aldo Moro 5, I-00185 Roma, Italy}

\author{Giorgio Milani}
\affiliation{Dipartimento di Fisica - Sapienza Universit\`{a} di Roma, P.le Aldo Moro 5, I-00185 Roma, Italy}

\author{Gonzalo Carvacho}
\affiliation{Dipartimento di Fisica - Sapienza Universit\`{a} di Roma, P.le Aldo Moro 5, I-00185 Roma, Italy}

\author{Rafael Chaves}
\email{rafael.chaves@ufrn.br}
\affiliation{International Institute of Physics, Federal University of Rio Grande do Norte, 59078-970, P. O. Box 1613, Natal, Brazil}

\author{Fabio Sciarrino}
\email{fabio.sciarrino@uniroma1.it}
\affiliation{Dipartimento di Fisica - Sapienza Universit\`{a} di Roma, P.le Aldo Moro 5, I-00185 Roma, Italy}

\begin{abstract}
\textbf{Since Bell's theorem, it is known that the concept of local realism fails to explain quantum phenomena. Indeed, the violation of a Bell inequality has become a synonym of the incompatibility of quantum theory with our classical notion of cause and effect. As recently discovered, however, the instrumental scenario --a tool of central importance in causal inference-- allows for signatures of nonclassicality that do not hinge on this paradigm. If, instead of relying on observational data only, we can also intervene in our experimental setup, quantum correlations can violate classical bounds on the causal influence even in scenarios where no violation of a Bell inequality is ever possible. That is, through interventions, we can witness the quantum behaviour of a system that would look classical otherwise. Using a photonic setup --faithfully implementing the instrumental causal structure and allowing to switch between the observational and interventional modes in a run to run basis-- we experimentally observe this new witness of nonclassicality for the first time. In parallel, we also test quantum bounds for the causal influence, showing that they provide a reliable tool for quantum causal modelling.}
\end{abstract}
\maketitle

\section{Introduction}
 
\begin{figure}[t!]
\center
\includegraphics[width=0.8\columnwidth]{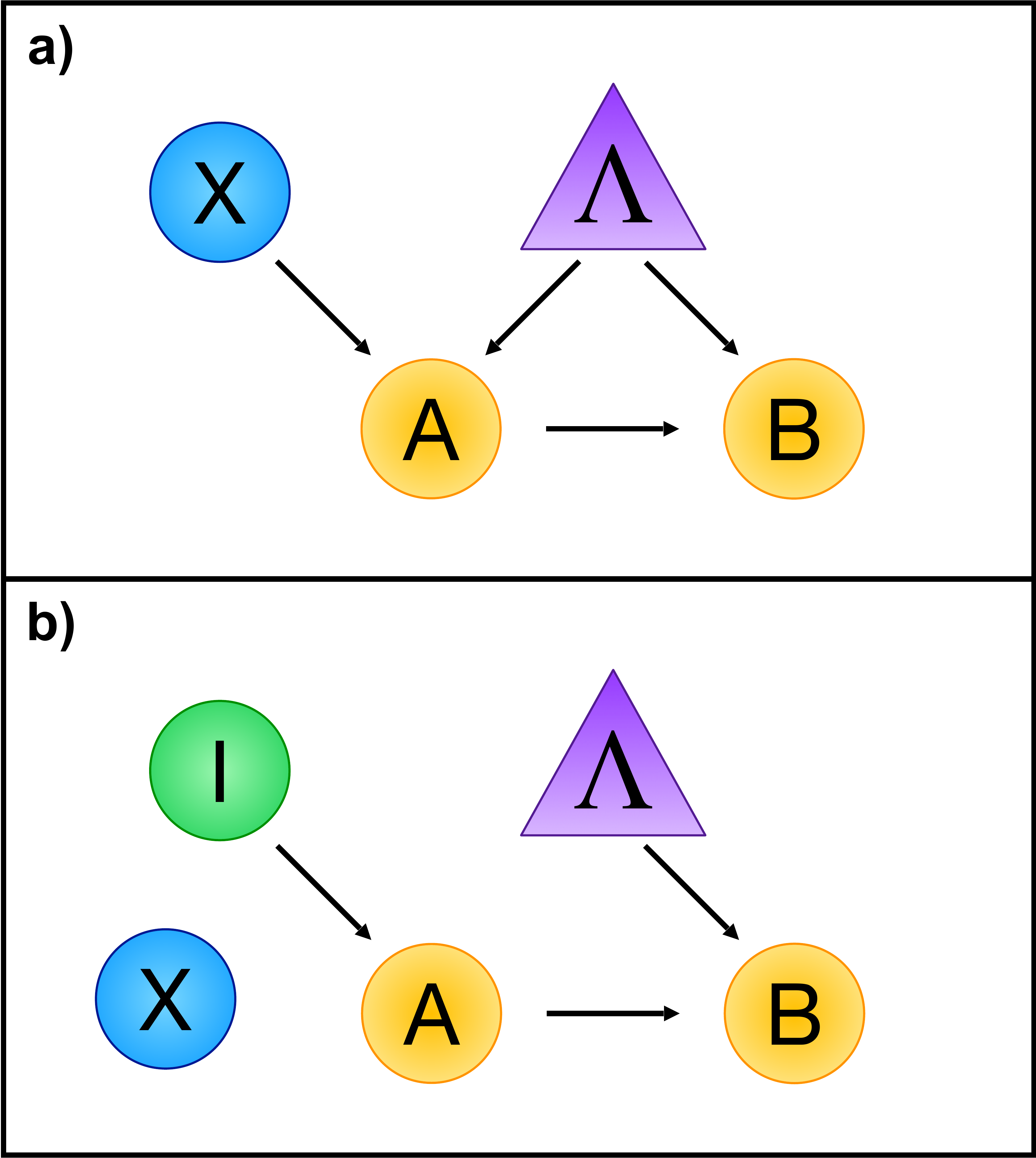}
\caption{ \textbf{Directed acyclic graph (DAG) of the instrumental scenario.}~ \textbf{a)} The instrumental scenario, where $X$ stands for the instrument, $A$ and $B$ are the variables between which causal influence is to be estimated and $\Lambda$ (that in the quantum case would be represented by a shared quantum state $\rho_{AB}$) represents any latent factor or common cause affecting them. \textbf{b)} Intervention in the instrumental scenario, where the independent variable $I$ forces the value of $A$, by cutting all the incoming arrows of $A$. In these graphs, circular nodes indicate observable variables, while latent variables are depicted as triangles.}
\label{fig:IDAG}
\end{figure}

The inference of cause-effect relationships from data is a keystone of any empirical science. Notwithstanding, distinguishing causation from correlations in practice is often a controversial matter. Without a direct intervention on the underlying mechanism generating the data, it might not be possible to distinguish between causation and confounding effects (common causes) \cite{pearlbook,pearl_2009}.  The simplest scenario in which that becomes possible is the instrumental causal model \cite{pearl_2013, bonet_2013}, shown in Fig.~\ref{fig:IDAG}a. That is achieved without any assumptions on the variables, apparatuses or physical mechanism involved, an approach known as device-independent in the context of quantum information \cite{acin_2007, pironio_2010, pironio_2016}. With the help of an instrumental variable $X$, the causal effect of a variable $A$ over another variable $B$ can be estimated without any interventions, a reason why such a tool has found use in a variety of fields \cite{wright_1928, goldberger_1972, bowden_1990, angrist_1995, greenland_2000, glymour_2001, morgan_2015, shipley_2016, peters_2017}. Nevertheless, to be applicable, one has to guarantee that the instrument satisfies a number of conditions implying testable constraints known as instrumental inequalities \cite{pearl1995testability}, the violation of which shows the inadequacy of the instrument being employed.

This causal inference framework, however, breaks down when quantum effects come into play. As recently discovered and experimentally demonstrated \cite{chaves_2018, himbeeck_2019, agresti_2019,agresti_2020}, with quantum entanglement acting as the common source, instrumental inequalities can be violated even by a perfect instrument. This not only shows that fundamental results in causality theory have to be reevaluated but also displays the value of moving beyond the paradigmatic Bell's theorem \cite{bell_1964, clauser_1969}, since considering different causal structures \cite{ fritz_2012, chaves_2016, rosset_2016, wolfe_2019, renou_2019, chaves_2018, poderini2020experimental} leads to new forms of nonclassical correlations and a broader understanding of the role of causality in quantum theory \cite{wood_2015,costa_2016,allen_2017,chaves_2021, carvacho_2019}. A remarkable feature of the instrumental scenario, the one we will focus on in this paper, is the fact that nonclassical behaviour can be witnessed without the need of violating a Bell inequality, something considered quintessential in standard scenarios \cite{scarani2019bell}. Considering the simplest instrumental scenario, with dichotomic variables only, it was theoretically shown that even though no Bell inequality can be violated by a quantum common source \cite{henson2014theory}, an entangled state can indeed violate the classical bounds for the causal effect of $A$ over $B$ \cite{gachechiladze_2020}. That is, even if the observed correlations admit a classical explanation (no Bell violation), they fail to do so if an intervention is performed.

In this article, we provide the first experimental demonstration of such a phenomenon. Exploiting interventions on a photonic platform equipped with an active feed-forward of information, implementing the causal scenarios in Fig.~\ref{fig:IDAG}, we detect a quantum signature in a setup that cannot violate any Bell inequality and thus would seem classical otherwise. More precisely, employing interventional data, we experimentally observe violations of the classical lower bounds for the causal influence between two variables, by producing several quantum states characterized by different degrees of entanglement. In addition, resorting to observational data, we test the quantum bound proposed in \cite{gachechiladze_2020}, showing its relevance for quantum causal modelling.  Our results offer a novel way to detect the presence of quantum correlations, without the need of violating Bell inequalities, which are often very demanding from the experimental perspective \cite{hensen2015loophole,shalm2015strong,giustina2015significant}.  We thus show that the incompatibility of quantum predictions with classical concepts can go beyond the paradigmatic Bell's theorem, opening a venue of research that might lead to new insights in quantum causality \cite{chaves2015information,costa_2016,allen_2017} and practical applications \cite{agresti_2020}.

\vspace{0.5cm}

\section{Measuring causal effects in classical and quantum physics}
The mantra in statistics that ``correlation does not imply causation'' subsumes the idea that correlations observed between variables $A$ and $B$ do not imply that one is the cause of the other, as a third, potentially unobservable, variable $\Lambda$ could be a source of the correlations. Direct causation and common source models generate the same set of possible probability distributions $p(a,b)$, making it impossible to distinguish both mechanisms from observational data alone unless further assumptions are imposed \cite{peters2017elements} or one is able to intervene in the system \cite{pearlbook, ringbauer2016experimental,ried2015quantum}. Considering the most general causal model,
\begin{equation}
    p(a,b)= \sum_{\lambda}p(\lambda)p(b\vert a,\lambda)p(a\vert \lambda),
\end{equation}
where the statistics of $B$ suffer both the influence of $A$ and the common source $\Lambda$, an intervention on $A$, denoted by $\mathrm{do}(a)$, erases all external influences on that variable, putting it under the experimenter's control and implying that
\begin{equation}
    p(\mathrm{do}(a),b)= \sum_{\lambda}p(\lambda)p(b\vert a,\lambda)p(\mathrm{do}(a)).
\end{equation}
That is, by intervening we effectively break all sources of confounding (see Fig. \ref{fig:IDAG}b). Interventions offer a natural way to quantify the causal influence of $A$ over $B$, for instance, via a measure called average causal effect (ACE) \cite{pearlbook,balke1997bounds} defined as
\begin{equation}
\label{eq:cACE}
    \mathrm{ACE} =  \max_{a,a',b} \vert p(b|\mathrm{do}(a))-p(b|\mathrm{do}(a')) \vert
\end{equation}
By this definition, if $\mathrm{ACE}$ is non-zero, we can be sure that $A$ has a direct causal influence over $B$.

Despite the fact that the intervention is a powerful tool for identifying causal relationships between two variables, it has a major drawback in that it can be difficult or even impossible to implement in practice. For this reason, one might rather consider the use of an extra variable, an instrumental variable $X$, which is in the full control of the experimenter, has a direct causal link to $A$ only, and is assumed to be independent of any confounding factors $\Lambda$. The corresponding causal structure can be represented by the directed acyclic graph (DAG) in Fig.~\ref{fig:IDAG}a, where each node represents a random variable and the directed edges encode their causal relations. Classically, such a causal model implies that the observed correlations $p(a,b \vert x)$ can be decomposed as
 
\begin{equation}
\label{eq:instrumental}
    p(a,b|x)=\sum_{\lambda}p(a|x,\lambda)p(b|a,\lambda)p(\lambda) \;,
\end{equation}
where $a$ and $b$ are the values assumed by the random variables $A$ and $B$, respectively, and $x$ is the value of the instrument $X$. 
The probabilities of $B$ upon an intervention on $A$ can then be calculated as
\begin{equation}
\label{eq:instrumental_do}
    p(b|\mathrm{do}(a))=\sum_{\lambda}p(b|a,\lambda)p(\lambda) \;,
\end{equation}
where the conditional probabilities $p(b|a,\lambda)$ and $p(\lambda)$ are the same as in Eq.~(\ref{eq:instrumental}). When the probabilities $p(b|\mathrm{do}(a))$ are calculated as in Eq.~(\ref{eq:instrumental_do}), i.e., when the experiment in question includes no quantum effects, we  refer to the average causal effect in Eq.(\ref{eq:cACE}) as cACE. 

Strikingly, as proven in Ref.~\cite{balke1997bounds}, the instrumental scenario  allows for estimating the strength of causal influence $\mathrm{cACE}$ lower bound as
\begin{equation}
\begin{split}
\centering
\label{eq:classbounds}
    &\mathrm{cACE}\ge  \mathrm{cACE_{LB}}=
    \\ &2p(0,0|0)+p(1,1|0)+
    p(0,1|1)+p(1,1|1)-2 \;.
\end{split}
\end{equation}
That is, simply relying on the observed data $p(a,b \vert x)$, we can estimate the effect of a possible intervention. Moreover, the estimation is achieved device-independently without resorting to the precise description of the system under study.

Let us give an example that demonstrates the significance of lower bounds on cACE. Consider the investigation of the impact of smoking on the development of cancer. In some study group, let $A$ stand for smoking/non-smoking and $B$ for cancer/no cancer.
The intervention on $A$ is not possible because it is simply unethical to force someone to smoke. However, we can introduce an instrumental variable $X$ which represents high/low taxation on tobacco and with high confidence has a direct effect only on $A$.  Then, by utilizing the lower bound in Eq.~\eqref{eq:classbounds} we can estimate the average causal effect of smoking on the risk of developing cancer. Note, however, that one has to guarantee that the used instrument complies with the instrumental decomposition in Eq.~\eqref{eq:instrumental}. Nicely, the causal assumptions entering in Eq.~\eqref{eq:instrumental} imply testable constraints called instrumental inequalities \cite{pearlbook,pearl1995testability}, whose violation signals the use of an inappropriate instrument. Hence, instrumental inequalities are for causal inference, what Bell inequalities are for Bell's theorem \cite{bell_1964}. From a causal perspective, Bell and instrumental inequalities are thus nothing else than classical constraints arising from imposing a causal structure to a given experiment.

\begin{figure}[t!] 
		\includegraphics[width=\columnwidth]{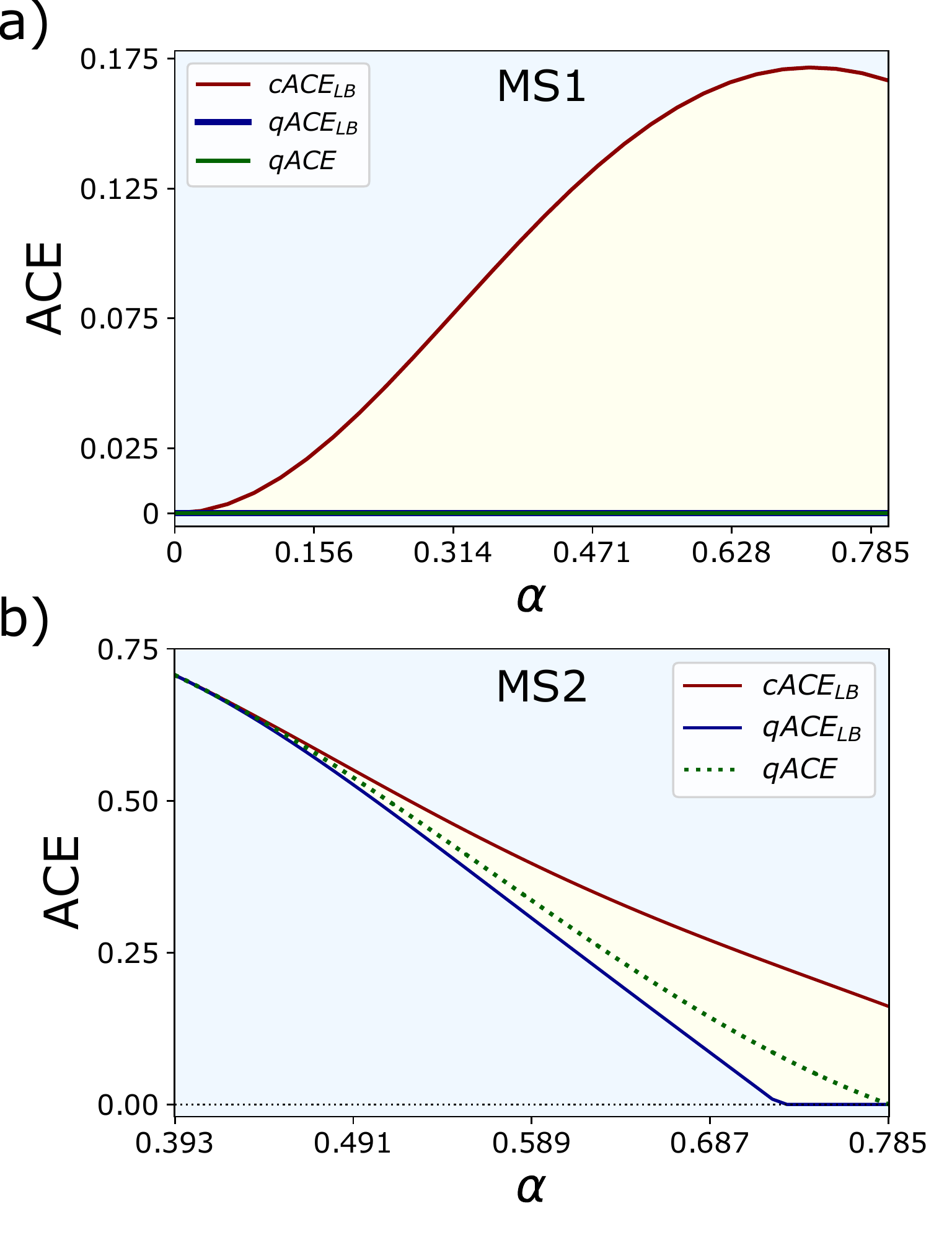}
		\caption{\textbf{Quantum and classical predictions for the average causal effect in the instrumental process.} 
		Within the instrumental process, $\mathrm{qACE}$ (green curve) can be lower than predicted by the classical theory of causality. Such a gap between the classical $\mathrm{cACE_{LB}}$ (red curve) and quantum  $\mathrm{qACE_{LB}}$ (blue curve) can emerge through suitable choices of the state shared by the parties and of the performed measurements. For the state in Eq.~\eqref{eq_state}, we report the quantum violation, i.e., $\mathrm{cACE}_{LB} > \mathrm{qACE}$, obtained by two measurement settings choices, MS1 (a) and MS2 (b), for a degree of entanglement $\alpha>0$ (see \textit{Methods}). The regions of interest where classical lower bounds are violated are depicted in yellow. The difference between the plots relies on the choice of measurement operators (see \textit{Methods}).
		\textbf{a)} Even though $\mathrm{qACE}$ is zero, a classical explanation requires a substantial amount of causal influence to explain the observed correlations.
		\textbf{b)} Example of observational correlations where the quantum bound becomes non-trivial and thus shows that causal influences can be estimated even in the presence of quantum common causes.} 
		\label{fig:regions}
\end{figure}

As one could expect from Bell's theorem, showing the incompatibility of quantum theory with local causality, it turns out that this classical framework is incompatible with quantum predictions. According to Born's rule, quantum correlations in the instrumental scenario are given by
\begin{equation}\label{eq:observedcorrelations}
    p(a,b|x)=\Tr[{(M_a^x\otimes N^a_b) \rho_{AB}}],
\end{equation}
where  the common source is a bipartite quantum state $\rho_{AB}$ and $M_a^x$ and $N_b^a$ are the operators describing the measurements on each subsystem. Note that $x$ is used to choose Alice's measurement setting and the outcome $a$ of Alice's measurement is used to determine Bob's measurement setting, accordingly. It is known that the quantum description is incompatible with the classical decomposition in Eq.~\eqref{eq:instrumental}. More precisely, if we impose the instrumental causal structure but allow the common source to be an entangled state, instrumental inequalities can be violated, if the instrument takes at least three possible values \cite{chaves_2018}. That is, something that classically would be seen as the violation of a causal assumption, in the quantum regime has to be understood as the presence of entanglement.

\begin{figure*}[t!]
\includegraphics[width=\textwidth]{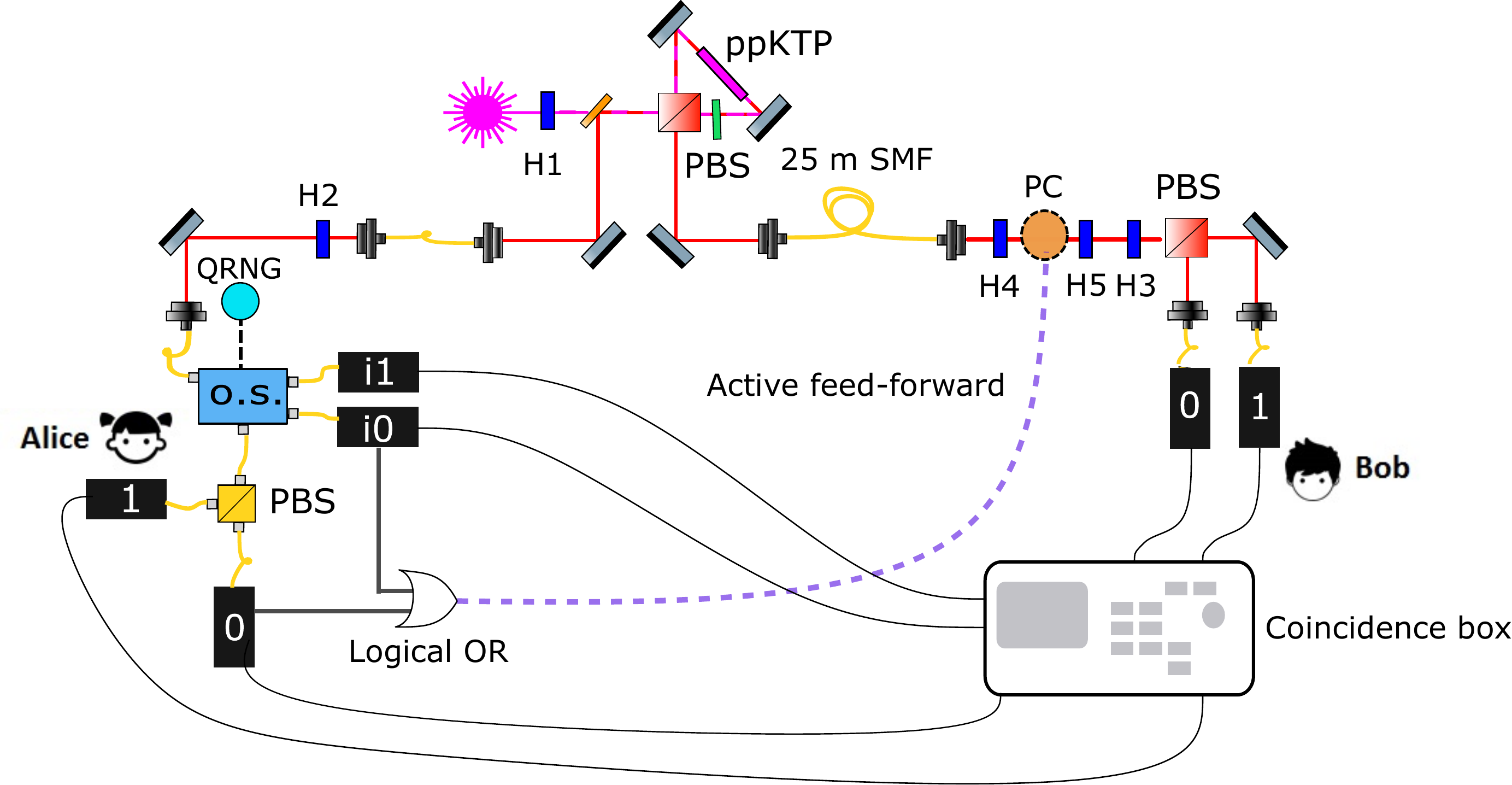}
\caption{ \textbf{Experimental apparatus}. A polarization-entangled photon pair is generated via spontaneous parametric down-conversion (SPDC) type II in a periodically poled titanyl phosphate, placed inside a Sagnac interferometer. The degree of entanglement generated can be selected by rotating the half-wave plate (HWP) $H1$.
On Alice's side, a quantum random number generator (QRNG) and an optical switcher (O.S. in the figure) select, for each experimental run, if an observation or an intervention is carried out. Alice's measurement basis is selected by rotating the HWP $H2$. Instead, if an intervention is performed, Alice's outcome is determined by the QRNG, by sending the photon either to detector $i1$ or $i0$. On Bob's side, a fast electro-optical device (Pockels Cell, PC) switches between the measurement bases, by inserting (or not) a $\pi$ phase between two orthogonal polarization states (selected by rotating $H4$ and $H5$). After that, a fixed HWP $H3$ is placed, followed by a bulk polarizing beam splitter (PBS) to perform the polarization projection. To implement the feed-forward of classical information between the measurement stations, Bob's photons is delayed
through a $25~m$ long single mode fiber, by $\sim~120~ns$.}
\label{fig:ExpSetupV1}
\end{figure*}

Interestingly, for the simplest instrumental scenario where all variables, including the instrument, are binary, even subtler quantum effects manifest themselves. The only class of instrumental inequalities in this scenario cannot be violated \cite{henson2014theory}, implying that all possible observed correlations $p(a,b \vert x)$  have a classical explanation. That, however, does not preclude non-classical effects for interventions, the latter defined in the quantum case as
\begin{equation}\label{eq:do_prob_quant}
   p(b|\mathrm{do}(a))=\Tr{[(\mathbbm{1} \otimes N_b^a) \rho_{AB}]}=\Tr{[N_b^{a} \rho_{B}]},
\end{equation}
where $\rho_B$ is the reduced state of Bob's system. Under an intervention the observed quantum average causal effect (qACE) is thus given by
\begin{equation}\label{eq:qACE}
    \mathrm{qACE}= \max_{a,a',b}(\Tr[(N_b^{a}-N_b^{a'})\rho_B]).
\end{equation}
In this quantum description, the classical bound in Eq.~\eqref{eq:classbounds} no longer holds and the qACE is rather lower bounded by \cite{gachechiladze_2020}
\begin{eqnarray}\label{eq:qACE_lb}
  \mathrm{qACE}_{LB}=\sum_{x=0,1}(p(0,0|x)+p(1,1|x))-\zeta-1,
 \end{eqnarray}
where $\zeta = \min_{\pm}\{\prod_{a=0,1}[1\pm\sum_{x=0,1}(-1)^x(p(a,0|x)-p(a,1|x))]\}^{\frac{1}{2}}$. It follows, that, for some probability distributions, i.e. for given states and observables, the following condition can occur:
\begin{equation}
\label{cqbounds}
   \mathrm{cACE} \geq \mathrm{cACE}_{LB} \geq \mathrm{qACE} \geq \mathrm{qACE}_{LB} \;,
\end{equation}
implying that the amount of quantum causal influence between $A$ and $B$ can be lower than the minimum required by any classical system. This proves that even if no Bell/instrumental inequality is violated, one can still witness non-classicality via interventions.

In the following, we show two instances of this non-classical behaviour, corresponding to correlations produced by a bipartite quantum state given by
\begin{equation}
    \ket{\psi(\alpha)} = \cos(\alpha)\ket{00} + \sin(\alpha)\ket{11},
    \label{eq_state}
\end{equation}
and two sets of different measurement settings, to which we will refer as MS1 and MS2.
In Fig.~\ref{fig:regions}, in particular, the bounds $\mathrm{cACE}_{LB}$ and $\mathrm{qACE}_{LB}$ are depicted by the red and blue curves, respectively, and the measured value of $\mathrm{qACE}$ (if an intervention is performed) is given by the green curve.
Such curves are functions of the parameter $\alpha$, which characterizes the entanglement of the state in Eq.~\eqref{eq_state}.
In further detail, the measurement setting  MS1, corresponding to Fig.~\ref{fig:regions}a, shows that, even if a quantification of the qACE leads to trivial values for every angle $\alpha$, nonetheless a classical explanation for the resulting correlations would require a non-zero amount of causal influences, as soon as $\alpha > 0$. This result, besides being a signature of a quantum behaviour, can also be interpreted as a quantum advantage in generating such correlations. Another interesting feature is that the maximum quantum violation is achieved for a non-maximally entangled state \cite{gachechiladze_2020}. 
On the other hand, MS2, corresponding to the results displayed in Fig.~\ref{fig:regions}b, witnesses that, in addition to a quantum violation of the cACE lower bound, also non-trivial amounts of quantum causal influences can be achieved. In other words, even in the presence of a quantum common source, we can put a non-trivial lower bound on the causal influence without the need of interventions.

In this work, our aim is to experimentally demonstrate the aforementioned predicted quantum violations, displayed in Fig.~\ref{fig:regions}. By doing so, we  experimentally show, for the first time, non-classical behaviours in a scenario where no standard quantum violation of a Bell/instrumental inequality is achievable.
A detailed description of the measurements belonging to the two settings can be found in the \textit{Experimental setup} and \textit{Methods} sections.
 
\begin{figure*}[tb]
\includegraphics[width=\textwidth]{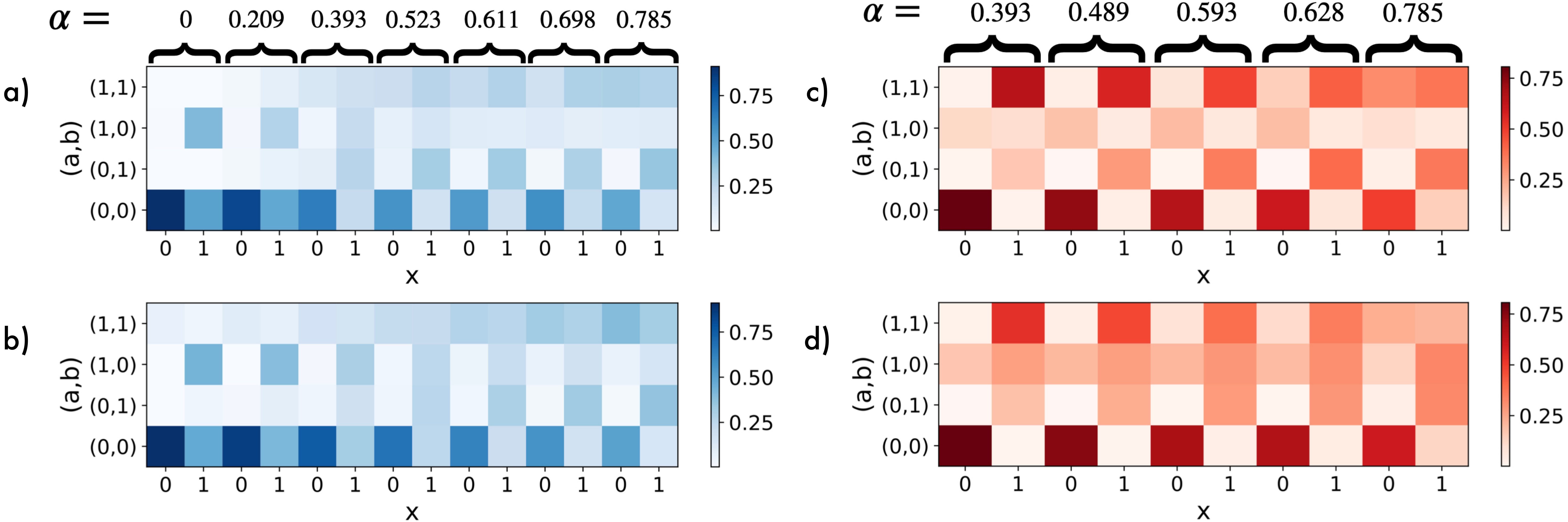}
\caption{ \textbf{Experimental and modelled $p(a,b|x)$  statistics.} The pair of columns of the colourplot refers to different amounts of entanglement ($\alpha$), see Eq.~\eqref{eq_state}, and the two columns report, respectively, the case of $x=0$ and $x=1$. The rows correspond to different $(a,b)$ tuples. \textbf{a)-b)} The reported values correspond to the experimental realization of the case of MS1, depicted in Fig.~\ref{fig:regions}a, and \textbf{c)-d)} to MS2, reported in Fig.~\ref{fig:regions}b. In both cases, in the top part, we report the experimental frequencies, while, in the bottom one, the probabilities given by a model of our apparatus, taking into account several sources of noise and imperfections (see Supplemental Material).
}
\label{fig:probs}
\end{figure*}

\section{Experimental Setup}

In order to test quantum violations of causal bounds as in Eq.~\eqref{cqbounds}, we need an experimental apparatus implementing the instrumental causal processes represented in Fig.~\ref{fig:IDAG}, allowing for the generation of both observational and interventional data. The causal structure in Fig.~\ref{fig:IDAG}a is used to observe correlations of the kind $p(a,b|x)$ and evaluate the classical and quantum lower bounds $\mathrm{cACE}_{LB}$ and $\mathrm{qACE}_{LB}$, through Eq.~\eqref{eq:classbounds} and Eq.~\eqref{eq:qACE_lb}, respectively. In turn, with the intervention illustrated in Fig.~\ref{fig:IDAG}b, the probabilities $p(b|\mathrm{do}(a))$ are retrieved, in order to evaluate the $\mathrm{qACE}$, as in Eq.~\eqref{eq:qACE}. To enforce that the observational data and the interventional one refer to the same experimental conditions, it is pivotal that, when changing between both configurations, the apparatus is maintained unaltered. Furthermore, to exclude time-dependent behaviours, it is crucial that at each experimental run we can decide randomly whether to implement the observational setup of Fig.~\ref{fig:IDAG}a or the interventional one of Fig.~\ref{fig:IDAG}b.

We achieve these conditions by exploiting the photonic platform depicted in Fig.~\ref{fig:ExpSetupV1}. Through a process of spontaneous parametric down-conversion (SPDC) in a periodically poled tytanil phosphate crystal within a Sagnac interferometer, we generate two-photon polarization entangled states. Considering the presence of both white and coloured noise in our quantum state, typical of SPDC quantum state sources  \cite{cabello_2005}, our experimental states are well modelled as (see \textit{Supplemental Material} for further details)
\begin{equation}
\begin{split}
    \rho_{noise} &= v | \psi^+ \rangle \langle \psi^+ | + \\
    &+ (1-v) \left(  \frac{\lambda}{2} (| \psi^+ \rangle \langle \psi^+ | + | \psi^- \rangle \langle \psi^- |) + \frac{1-\lambda}{4} \mathbb{I} \right)
\end{split}
\label{eq:noise}
\end{equation}
where $| \psi^{\pm} \rangle = \cos(\alpha) |00 \rangle \pm  \sin(\alpha) |11 \rangle$ and $\ket{0}$ and $\ket{1}$ are encoded in the horizontal and vertical photon polarization. The $\alpha$ parameter is set by rotating the source half-wave plate (HWP) H1 (see Fig.\ref{fig:ExpSetupV1}) by $\alpha/2$. 

Since the instrumental causal structure implies a direct causal influence from Alice to Bob, we implement an active exchange of information between the two parties, by employing a fast electro-optical device (Pockels Cell, PC). When such a device is triggered by the application of a high voltage, it behaves like a wave retarder, inserting a $\pi$ phase shift between two polarizations forming an orthogonal basis. Instead, when not triggered, it performs the identity operator. 

To switch between the causal processes in Fig.~\ref{fig:IDAG}a and Fig.~\ref{fig:IDAG}b, an optical switcher is put on Alice's side. Such a switcher is controlled by a quantum random number generator (QRNG) and selects on which out of three paths Alice's photons are sent. In particular, the QRNG generates two bits $q_1$ and $q_2$. If $q_1=0$, photons are sent to an in-fiber polarizing beam splitter (PBS), whose output modes are connected to two detectors $0$ and $1$, corresponding to Alice's possible outcomes, implementing a regular measurement generating the observational data $p(a,b \vert x)$. The measurement basis is selected through the half-wave plate $H2$, rotated by an angle $\eta_x$. In this way, Alice's measurement will correspond to a projective measurement on the x-z plane of the Bloch sphere, given by
\begin{equation}
    M(\eta_x)=\cos(\eta_x)\sigma_z+ \sin(\eta_x)\sigma_x
\end{equation}
Instead, if $q_1=1$, no measurement is performed and an intervention is carried out. In particular, if $q_2=0$, Alice's outcome is forced to be $0$ and photons are sent to the detector $i0$. On the contrary, if $q_2=1$, Alice's outcome is forced to be $1$ and photons are sent to detector $i1$. Hence, the QRNG represents the independent variable $I$ belonging to the causal structure in Fig.~\ref{fig:IDAG}b.

On the other side, Bob's measurement station is composed of a half-wave plate (H3), rotated by $\phi_1/4$ and preceded by the PC, which is put between two half-wave plates (H4 and H5), rotated by $\theta/2$. In this way, when triggered, the cell inserts a $\pi$ shift between the two orthogonal polarization states $\cos(\theta)\ket{0}+\sin(\theta)\ket{1}$ and $-\sin(\theta)\ket{0}+\cos(\theta)\ket{1}$. At the end, a PBS performs a projective measurement on x-z plane given by
\begin{equation}
    N(\phi_a)=\cos(\phi_a)\sigma_z+ \sin(\phi_a)\sigma_x.
\end{equation}

In order to reproduce both causal scenarios in Fig.\ref{fig:IDAG} properly, the PC needs to be activated when a regular measurement is performed and Alice's outcome is $0$, as well as when intervention $\mathrm{do}(A=0)$ is made. Hence, the electronic signals produced by detectors $0$ and $i0$ are inputted to a device performing a logic OR operation. The OR signal is sent to the PC driver, selecting Bob's measurement basis and thus switching from $\phi_1$ to $\phi_0$, while a copy of the signal from the detector is sent to a coincidence counter.
To achieve the feed-forward of information between Alice and Bob (the causal arrow between their measurement outputs), Bob's photon is sent through a $25~\mathrm{m}$ long single mode fiber, corresponding to a delay of $\sim 120~\mathrm{ns}$ .
Details on the angles of the measurement operators can be found in the \textit{Methods} section.

\begin{figure*}[t!]
\includegraphics[width=\textwidth]{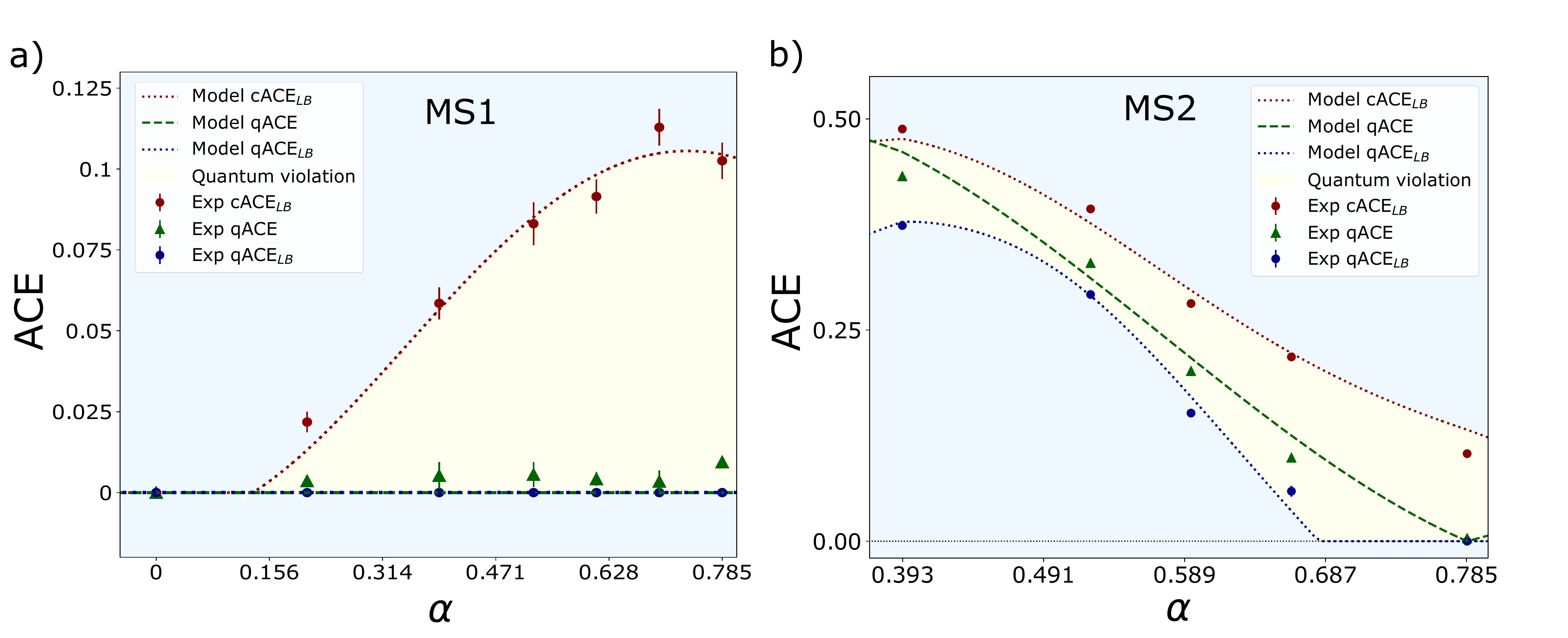}
\caption{ \textbf{Experimental average causal influence results.}
The two plots report the experimental results corresponding to the parties sharing the state in Eq.~\eqref{eq_state} and performing, respectively, the measurement setting MS1 (a) and MS2 (b) (see Methods). Both cases were experimentally implemented through the apparatus depicted in Fig.~\ref{fig:ExpSetupV1}. The red points correspond to the experimental values of the $\mathrm{cACE}_{LB}$, while the blue ones correspond to those of the $\mathrm{qACE}_{LB}$, evaluated through the observable probabilities $p(a,b|x)$, for different amounts of entanglement $\alpha$ (see Eq.~\eqref{eq_state}). The yellow colour highlights the regions where $\mathrm{cACE}_{LB} > \mathrm{qACE}_{LB}$ and a quantum violation is possibly achievable.
The green triangles represent the measure of quantum ACE $\mathrm{qACE}$, obtained via the probabilities $p(b|\mathrm{do}(a))$ by intervening on $A$. The bars represent the uncertainty on all the values, amounting to one standard deviation, which were computed through Monte Carlo simulations. Let us note that, for MS1, the experimental points corresponding to all $qACE_{LB}$ values and to the $cACE_{LB}$ for $\alpha=0$, from Eqs.~\eqref{eq:classbounds} and \eqref{eq:qACE_lb}, amount to negative, and therefore trivial, values, so, in the graph, we set those points to $0$. The same occurs, in case \textbf{b}, for the $qACE_{LB}$ in $\alpha=0.785$. The confidence level with which we can conclude that this operation is fair can be quantified by the distance of the experimentally obtained values from $0$, through Gaussian standard tables. In detail, the lowest distance amounts to more than $10$ standard deviations. We report all of the experimental values in the Supplementary material. The dotted curves are the theoretical benchmarks, taking into account the experimental imperfections of our apparatus.  \textbf{a)} Alice's and Bob's operators, constituting MS1, are chosen to maximize the quantum violation of the classical causal bound $\mathrm{cACE}_{LB}$. \textbf{b)} Alice's and Bob's operators, constituting MS2, are chosen to show non-trivial bounds on both the classical and quantum ACE.}
\label{fig:curva_punti}
\end{figure*}

\section{Results}

For different values of $\alpha$ in the entangled state of Eq.~\eqref{eq_state}, we collected the observational data $p(a,b|x)$ as well as the interventional data $p(b\vert \mathrm{do}(a))$. With the former, we computed the lower bounds for the causal influence $\mathrm{cACE}_{LB}$ and $\mathrm{qACE}_{LB}$, prescribed by Eq.~\eqref{eq:classbounds} and Eq.~\eqref{eq:qACE_lb}, respectively. With the second set of data, we quantified the actual value of the $\mathrm{qACE}$ as described in Eq.~\eqref{eq:qACE}. As previously mentioned, our aim is to experimentally reproduce the curves in Fig.~\ref{fig:regions}, brought by the measurement settings MS1 and MS2 on a state of the form reported in Eq.~\eqref{eq_state}. In particular, MS1 maximizes the quantum violation of the classical causal bound Eq.~\eqref{eq:cACE} for given amounts of entanglement in the source (Fig.~\ref{fig:regions}a). In turn, MS2 (Fig.~\ref{fig:regions}b) shows cases where the quantum bound leads to non-trivial (larger than zero) lower bounds on the quantum ACE.

To benchmark our experimental results, we not only take into account the noisy state of Eq.~\eqref{eq:noise}, but also other imperfections within the apparatus, e.g., the non-perfect $\pi$ phase inserted by the PC, the different efficiencies of the detectors and possible imperfections in the wave plate rotation angles. A comparison between modelled and experimental results of the observational data $p(a,b \vert x)$, for the two scenarios, is presented in Fig.~\ref{fig:probs}, showing that those probabilities are very well described by our modelling of the experiment.

In turn, Fig.~\ref{fig:curva_punti} shows in details the quantum violation of the classical causal bound in Eq.~\eqref{eq:cACE}. More precisely, by performing interventions, we can show that the experimentally measured quantum causal effect $\mathrm{qACE}$ (green triangles) violates the classical lower bound $\mathrm{cACE}_{LB}$ (red points). In particular, Fig.~\ref{fig:curva_punti}a is obtained performing the measurements belonging to MS1 and it shows that the maximal violation of the classical bound does not correspond to a maximally entangled state. In turn, Fig.~\ref{fig:curva_punti}b, corresponding to the set MS2, adds a complementary aspect, showing that $\mathrm{qACE}_{LB}$ (blue dots) can provide non-trivial and fairly tight lower bound to the estimation of causal influence even in the presence of a source of quantum correlations.

\section{Discussion}

In this work, we have experimentally shown that the instrumental scenario leads to a new kind of nonclassical correlation. Differently from Bell's theorem, we have a temporal correlation scenario where Bob's measurement input is defined by Alice's measurement output. Within this context, we consider the simplest case where all observable variables are binary: a situation where it is known that no Bell inequality can be violated \cite{henson2014theory}. This implies that all of the observed correlations in our experiment do have a classical explanation. However, by implementing interventions in our setup and quantifying the causal influence of Alice's outcome $A$ on Bob's output $B$, the quantum nature of our setup is revealed. Indeed, our interventional data violates a classical causal bound for the average causal effect of $A$ over $B$, which constitutes a novel signature of nonclassicality that only recently has been theoretically discovered \cite{gachechiladze_2020}.

Our photonic setup faithfully implements the temporal causal structure underlying the instrumental scenario. Furthermore, the use of a quantum random number generator together with an optical switcher permit us to decide, on a run to run basis, whether to perform an observational or interventional measurement. This is crucial to ensure that the observational data used to compute the classical and quantum lower bounds to ACE, as well as the interventional data used to compute ACE directly, do refer to the same experimental conditions. The versatility of the setup to generate quantum states with different degrees of entanglement and the ability to measure on different basis allow us to probe the quantification of quantum causality in different regimes. For instance, focusing on the optimum quantum violation of causal bounds, we were able to experimentally show that a non-maximally entangled state leads to maximum nonclassicality. In turn, by changing the measurement operators, we not only showed nonclassical bahaviour but also experimentally proved that the quantum causal bounds derived in \cite{gachechiladze_2020} provide a fairly good estimation of quantum causal influences.

The main contribution of this work is to experimentally prove, for the first time, a new kind of nonclassical behaviour that do not hinge on the paradigm of a Bell inequality violation. It is worth noticing that, similar to the Bell scenario, our conclusions are reached in a device independent setting \cite{pironio_2016}. Indeed, we are imposing the causal structures of Fig.~\ref{fig:IDAG} to the experiment and performing the intervention on a classical variable (Alice's outcome), so that the whole analysis only rely on the observed input/output correlations and does not need to assume anything on measurement and state preparation devices employed by Alice and Bob. We are imposing the causal structure to the experiment and the whole analysis only rely on the observed input/output correlations and does not need to assume anything on measurement and state preparation devices. Our results open a new venue on the foundations of quantum theory and, in particular, on the role of causality on quantum effects. In particular, notice that the instrumental causal structure underlies the remote state preparation \cite{bennett_2001} and teleportation \cite{bennett1993teleporting} protocols, hinting at the possibility of revisiting paradigmatic quantum tasks from the causal perspective. On the more applied side, it is known that the instrumental scenario can also be employed in cryptography protocols \cite{agresti_2020}, but the role of causal effects and interventions on such protocols remains, to our knowledge, completely unexplored. We hope our finding might trigger future developments along these and other promising lines of research.

\section{acknowledgments}
This work was supported by the PRIN-MIUR (Italy) Grant QUSHIP (Taming complexity with quantum strategies: a hybrid integrated photonics approach) Id. 2017SRNBRK, by the John Templeton Foundation via the grant Q-CAUSAL No 61084
 and via The Quantum Information Structure of Spacetime (QISS) Project (qiss.fr) (the opinions expressed in this publication are those of the author(s) and do not necessarily reflect the views of the John Templeton Foundation) 
Grant Agreement No. 61466, by Progetti per Avvio alla Ricerca 2018-2019 offered by Sapienza Universit\`{a} di Roma and is also supported by the Serrapilheira Institute (grant number Serra-1708- 15763).
 RC also acknowledge the Brazilian National Council for Scientific and Technological Development (CNPq) via the National Institute for Science and Technology on Quantum Information (INCT-IQ) and Grants No. 406574/2018-9 and 307295/2020-6, the Brazilian agencies MCTIC and MEC. N.M.~acknowledges the support by the Foundation for Polish Science (IRAP project, ICTQT, contract no.~2018/MAB/5, co-financed by EU within Smart Growth Operational Programme) and the Deutsche Forschungsgemeinschaft (DFG, German Research Foundation) via the Emmy Noether grant 441423094. M.G.~is funded by the Deutsche Forschungsgemeinschaft (DFG, German Research Foundation) under Germany's Excellence Strategy – Cluster of Excellence Matter and Light for Quantum Computing (ML4Q) EXC 2004/1 – 390534769.

\subsection{Competing financial interests}
The authors declare no competing financial interest.

\bigskip
\section{Methods}
\subsection{Experimental Details}
Photon pairs are generated in a parametric down conversion source, composed by a $20~mm$-thick periodically poled KTP (ppKTP) crystal inside a Sagnac interferometer. 
The source employs a continuous wave diode laser with wavelength of $\lambda=404$~nm. Photons generated are filtered in wavelength and spatial mode by using narrow band interference filters and single-mode fibers, respectively. 
The crystal used to implement active feed-forward is a $\mathrm{LiNbO}_{3}$ high-voltage micro PC made by Shangai Institute of Ceramics with $\sim 90~\mathrm{ns}$ risetime and a fast electronic circuit transforming each Si-avalanche photodetection signal into a calibrated fast pulse in the kV range needed to activate the PC. To achieve the active feed-forward of information, the photon sent to Bob's station needs to be delayed, thus allowing the measurement on the first qubit to be performed. The amount of delay was evaluated considering the detectors' response time, the velocity of the signal transmission through a single mode fiber, whose refraction index $\sim 1.45$ and the activation time of the PC. Therefore, we have used a fiber $25~\mathrm{m}$ long, coupled at the end into a single mode fiber that allows a delay of $\sim 120~\mathrm{ns}$ of the second photon with respect to the first. The voltage applied to the Cell, in order to insert a $\pi$ shift between the two polarizations was of $1350~\mathrm{V}$.
The QRNG which controls the optical switcher on Alice's station is an IdQuantique product (model: Quantis-USB-4M).

The curve in Fig.~\ref{fig:regions}a corresponds to the measurement setting MS1, which requires $\phi_1=-\phi_0$, $\eta_0=arctg(\frac{\sin(2\alpha) \sin(\phi_0)}{\cos(2\alpha)+3 \cos(\phi_0)})$, $\eta_1=-\frac{\pi}{2}$ and $\phi_0$ is chosen to maximize the difference between the classical predictions for the $\mathrm{cACE}_{LB}$ and the $\mathrm{qACE}$ (see \cite{gachechiladze_2020} and Supplemental Material note 2). To switch from $N(\phi_0)$ to $N(\phi_1=-\phi_0)$, the PC and the HWPs $H4$ and $H5$ are in their optical axis and $H3$ is rotated of $\phi_1/4$ (see Fig.~\ref{fig:ExpSetupV1} and Supplementary Material note 2 for further details). The voltage applied to the PC, in order to insert a phase of $\pi$, is of $1350V$.

For the measurement setting MS2 (Fig.~\ref{fig:regions}b), instead, the measurement parameters are the following: $\eta_0=3(\alpha-\frac{\pi}{8})$, $\eta_1=\pi$, $\phi_0=2(\alpha-\frac{\pi}{8})$ and $\phi_1=\pi-3(\alpha-\frac{\pi}{8})$. In order to switch from $N(\phi_1)$ to $N(\phi_0)$, the PC is kept in its optical axis, while $H4$ and $H5$ are rotated of $\theta/2$, depending on $\alpha$, while $H3$ is rotated of $+\phi_1/4$ (see Fig.~\ref{fig:ExpSetupV1} and Supplementary Material note 2 for further details). Also in this case, the phase inserted by the PC is $\pi$ and the high voltage applied amounts to $1350V$.

\subsection{Data availability}
The data that support the plots within this paper and other findings of this study are available from the corresponding author upon request.

%

\end{document}


\title{Supplementary Information}

\author{Iris Agresti}
\affiliation{Dipartimento di Fisica - Sapienza Universit\`{a} di Roma, P.le Aldo Moro 5, I-00185 Roma, Italy}

\author{Davide Poderini}
\affiliation{Dipartimento di Fisica - Sapienza Universit\`{a} di Roma, P.le Aldo Moro 5, I-00185 Roma, Italy}

\author{Beatrice Polacchi}
\affiliation{Dipartimento di Fisica - Sapienza Universit\`{a} di Roma, P.le Aldo Moro 5, I-00185 Roma, Italy}

\author{Nikolai Miklin}
\affiliation{International Centre for Theory of Quantum Technologies (ICTQT), University of Gdansk, 80-308 Gdansk, Poland}
\affiliation{Heinrich Heine University D{\"u}sseldorf, Universit{\"a}tsstra{\ss}e 1, 40225 D{\"u}sseldorf, Germany}

\author{Mariami Gachechiladze}
\affiliation{Institute for Theoretical Physics, University of Cologne, 50937 Cologne, Germany}

\author{Alessia Suprano}
\affiliation{Dipartimento di Fisica - Sapienza Universit\`{a} di Roma, P.le Aldo Moro 5, I-00185 Roma, Italy}

\author{Emanuele Polino}
\affiliation{Dipartimento di Fisica - Sapienza Universit\`{a} di Roma, P.le Aldo Moro 5, I-00185 Roma, Italy}

\author{Giorgio Milani}
\affiliation{Dipartimento di Fisica - Sapienza Universit\`{a} di Roma, P.le Aldo Moro 5, I-00185 Roma, Italy}

\author{Gonzalo Carvacho}
\affiliation{Dipartimento di Fisica - Sapienza Universit\`{a} di Roma, P.le Aldo Moro 5, I-00185 Roma, Italy}

\author{Rafael Chaves}
\email{rafael.chaves@ufrn.br}
\affiliation{International Institute of Physics, Federal University of Rio Grande do Norte, 59078-970, P. O. Box 1613, Natal, Brazil}

\author{Fabio Sciarrino}
\email{fabio.sciarrino@uniroma1.it}
\affiliation{Dipartimento di Fisica - Sapienza Universit\`{a} di Roma, P.le Aldo Moro 5, I-00185 Roma, Italy}

\maketitle
\section{Noise model}

In this section, we describe the noise model which best fits our experimental data, reported in the curves shown in Fig.~5 of the main text.
In detail, we considered a noise model, including a fraction of white noise (the visibility $v$) and a fraction of colored noise ($\lambda$), which is typical of SPDC sources \cite{cabello_2005}.

White noise is the result of a partial isotropic depolarization of the state. Indicating the 2-qubit ideal pure state as  $\rho$, with  $ \rho = | \psi \rangle \langle \psi |$, the state affected by white noise is then given by
\begin{equation}
    \rho_{white} = v' \rho + \frac{(1-v')}{4} \mathbb{I}
\end{equation}

On the other hand, colored noise is an anisotropic depolarization, along a preferred direction.
Its action on the state $| \psi^+ \rangle \langle \psi^+ |$, with $ | \psi^{\pm} \rangle = \cos{(\alpha)} | 00 \rangle \pm \sin{(\alpha)} | 11 \rangle$, quantified by the parameter $\lambda'$, is described as
\begin{equation}
    \rho_{col} = \lambda' | \psi^+ \rangle \langle \psi^+ | + \frac{(1-\lambda')}{2}\left( | \psi^+ \rangle \langle \psi^+ | + | \psi^- \rangle \langle \psi^- |  \right).
\end{equation}

Therefore, our overall noisy state is modelled combining these different contributions in a normalized form given by
\begin{equation}
\begin{split}
    \rho_{noisy} &= v~|\psi^+ \rangle \langle \psi^+ |+ (1-v)~ \left( ~ \frac{\lambda}{2}~ (|\psi^{+} \rangle \langle \psi^+ | + |\psi^- \rangle \langle \psi^- |)  + \frac{(1-\lambda)}{4} \mathbb{I}  \right),
\end{split}
\label{eq:noise}
\end{equation}
where the visibilities have been redefined such that $v=1$ yields the pure state, while $\lambda $ refers to the fraction of colored noise with respect to the total visibility.

The parameters $v$ and $\lambda$ that best model our experimental state are
\begin{equation*}
    \begin{split}
    v &= 0.81, \\
    \lambda &= 0.93. \\
    \end{split}
\end{equation*}

An additional source of experimental noise that we considered, in order to predict the expected values for the cACE and qACE reported in Fig.~5 of the main text, is the non perfect phase $\delta$ introduced by the electro-optical device (Pockels cell, PC) employed in Bob's measurement station.
 
For our experiment, we require that $\delta=\pi$, so that, when triggered by high voltage, the action of such a device is identical of that of a half-wave plate (HWP). Otherwise, it performs the identity operation. However, the largest effective phase inserted by the PC (for a voltage of $1350V$) oscillated between $0.7 \pi$ and $0.8 \pi$. 

Indeed, the $\delta$ value which best models our experimental data equals $0.802 \pi$ for the curve in Fig.~5a of the main text and $0.716\pi$ for that in Fig.~5b.

Moreover, we also took into account minor noise sources such as the non perfect rotation of the adopted HWPs and the non perfect efficiencies of the detectors.

\section{Optimization and implementation of the operators}

In this supplementary section, we report the optimal operators for Alice and Bob, corresponding to the measurement settings $MS1$ and $MS2$, given the degree of entanglement $\alpha$, to reproduce, respectively, the curve in Fig.~2a and Fig.~2b of the main text.

We define the operator respectively measured by Alice and Bob using the notation \cite{gachechiladze_2020}
\begin{equation}
    M(\theta_x) := M^x = M_0^x - M^x_1, \quad N(\phi_a) := N^a = N^a_0 - N^a_1, \quad \{ x, a \} \in \{ 0,1\}, 
    \label{eq:operators}
\end{equation}
where, $M^x$ and $N^a$ are given by
\begin{equation}
    M^x = \cos{(\theta_x)} \sigma_z + \sin{(\theta_x)} \sigma_x ,
    \label{opm}
\end{equation}
and 
\begin{equation}
    N^a = \cos{(\phi_a)} \sigma_z + \sin{(\phi_a)} \sigma_x .
    \label{opn}
\end{equation}

Thus, $\mathrm{qACE}$ can be expressed as
\begin{equation}
    \mathrm{qACE}_{A \rightarrow B} = \max_b \{ tr [\mathbb{I} \otimes (N^0_b - N^1_b) | \psi \rangle \langle \psi |] \} = \frac{1}{2} | \langle  \mathbb{I} \otimes N(\phi_0) - N(\phi_1)  \rangle |.
\end{equation}

Without loss of generality, we can assume $\langle  \mathbb{I} \otimes N(\phi_0) - N(\phi_1)  \rangle \geq 0 $. If this is not the case, one can always consider the scenario where Alice relabels the measurement outcomes and the subsequent derivations would follow accordingly.

Following Eq.~(5) of the main text, the classical lower bound on the ACE, $\mathrm{cACE}_{A \rightarrow B}^*$, is defined as
\begin{equation}
    cACE_{A \rightarrow B}^* = 2 \langle M_0^0 \otimes N^0_0 \rangle + \langle M_1^0 \otimes N_1^1 \rangle + \langle M_0^1 \otimes N_1^0 \rangle + \langle M_1^1 \otimes N_1^1 \rangle - 2.
\end{equation}

This quantity can be rewritten, after some simplifications, as
\begin{equation}
    \mathrm{cACE}_{A \rightarrow B}^*  = \frac{1}{4} (-3 + \langle \mathbb{I} \otimes N(\phi_0) \rangle + \langle M(\theta_0) \otimes \mathbb{I} \rangle -2 \langle \mathbb{I} \otimes N(\phi_1) \rangle + f(\theta_0, \theta_1, \phi_0, \phi_1, \alpha ) ).
\end{equation}

The expected quantum violation of the causal bound, namely the quantity $cACE_{A \rightarrow B}^* - qACE_{A \rightarrow B}$, amounts to
\begin{equation}
    \mathrm{cACE}_{A \rightarrow B}^* - \mathrm{qACE}_{A \rightarrow B} = \frac{1}{4} \left( -3- \langle \mathbb{I} \otimes N(\phi_0) \rangle + \langle M(\theta_0) \otimes \mathbb{I} \rangle  + f(\theta_0, \theta_1, \phi_0, \phi_1, \alpha ) \right).
\end{equation}

It can be numerically shown that this quantity is maximal when imposing $\phi_1 = -\phi_0$ and $\theta_1 = - \frac{\pi}{2}$. In this case, the following equality holds for function $f$: $f(\theta_0, \theta_1, \phi_0, \phi_1, \alpha ) = 3\cos{(\theta_0)}\cos{(\phi_0)} + \sin{(2\alpha)}\sin{(\phi_0)}( 2 + \sin{(\theta_0)})$. This choice of angles $\{ \phi_0, \phi_1, \theta_1 \}$ characterizes the operators which reproduce the curve in Fig.~2a.
Moreover, let us note that the assumption $\phi_1 = -\phi_0$ ensures that $\mathrm{qACE} = 0$. The parameter $\theta_0$ can be optimized analytically such that it maximizes the violation, yielding
\begin{equation}
    \theta_0 = \text{arccot} \left( \frac{\cos{(2\alpha)} + 3\cos{(\phi_0)}}{\sin{(2\alpha)} \sin{(\phi_0)}}  \right)
\end{equation}.
The parameter $\phi_0$, the only parameter left undetermined, can be numerically optimized.
These operators are showed to yield the maximal violation extent, amounting to $3-2\sqrt{2}$, with $2\alpha = \arctan \left( \frac{1}{\sqrt{3 \sqrt{2} + 2}} \right) + \arctan \left( \sqrt{\frac{1}{2} \left( 3\sqrt{2} + 2 \right)}  \right) $ and $\phi_0 = \arctan \left( \frac{2}{\sqrt{3 \sqrt{2}} + 2} \right)$.

To reproduce the curve in Fig.~2b of the main text, we let the entanglement degree $\alpha$ range in the interval $\alpha \in [\pi/8, \pi/4]$, while the angles of Alice and Bob's operators are given by
\begin{equation}
    \begin{split}
        \theta_0 &= 3 \left( \alpha - \frac{\pi}{8} \right), \\
        \theta_1 &= \pi, \\
        \phi_0 &= 2 \left( \alpha - \frac{\pi}{8} \right), \\
        \phi_1 &= \pi - 3 \left( \alpha - \frac{\pi}{8} \right).
    \end{split}
    \label{ops2}
\end{equation}

In order to experimentally implement Bob's operators for $MS1$, we adopted a PC followed by a fixed HWP. The latter is rotated of $\frac{\phi_1}{4}$, so that, when the PC is not triggered and behaves like an identity operator, Bob performs the projective measurement onto the eigenstates of $N^1$, i.e. $\cos(\phi_1)\sigma_z+\sin(\phi_1)\sigma_x$. Then, in order to switch to $N^0$, i.e. $\cos(\phi_0)\sigma_z+\sin(\phi_0)\sigma_x$, with $\phi_0=- \phi_1$, when the Pockels cell is triggered, we require the PC to be in its optical axis and to insert a $\pi$ shift between the horizontal and vertical polarization states.
Indeed, the Pockels cell has the Jones matrix given by
\begin{equation}
P(\delta)=
    \begin{pmatrix}
    1 & 0 \\
    0 & e^{i\delta}
    \end{pmatrix},
\end{equation}
which, when $\delta=\pi$, is equal to $\sigma_z$. Hence, the combined action of the PC and the fixed HWP ammounts to
\begin{equation}
\sigma_z ~(\cos(\phi_1)\sigma_z+\sin(\phi_1)\sigma_x)~ \sigma_z=\cos(\phi_1)\sigma_z-\sin(\phi_1)\sigma_x=\cos(\phi_0)\sigma_z+\sin(\phi_0)\sigma_x=N^0.
\end{equation}

In order to experimentally implement Bob's operators for $MS2$ (see Eq.~\eqref{ops2}), we keep $\delta=\pi$ and the HWP fixed to $\frac{\phi_1}{4}$, but we change the polarization orthogonal states between which the PC inserts the phase, by rotating the PC of an angle $\eta$.
In this way, the combined action of the rotated PC and the fixed HWP is
\begin{equation}
    N(\eta)=R(\eta)P(\pi)R(-\eta)N^1R(\eta)P^\dag(\pi)R(-\eta)=R(\eta)\sigma_z R(-\eta)N^1R(\eta)\sigma_z R(-\eta),
\label{n_theta}
\end{equation}
where $R(\eta)$ is the rotation matrix 
\begin{equation}
R(\eta)=
    \begin{pmatrix}
    \cos(\eta) & -\sin(\eta) \\
    \sin(\eta)& \cos(\eta)
    \end{pmatrix}.
\end{equation}
At this point, in order to select the proper angle $\eta$ for each $\alpha$, the following system of equations is solved:
\begin{equation}
\begin{cases}
~Tr(M^0_0 \otimes N^0_0 \rho(\alpha))= Tr(M^0_0 \otimes N(\eta)_0 \rho(\alpha)),\\
~Tr(M^0_1 \otimes N^0_0 \rho(\alpha))= Tr(M^0_1 \otimes N(\eta)_0 \rho(\alpha)),\\
~Tr(M^0_0 \otimes N^0_1 \rho(\alpha))= Tr(M^0_0 \otimes N(\eta)_1 \rho(\alpha)),\\
~Tr(M^0_1 \otimes N^0_1 \rho(\alpha))= Tr(M^0_1 \otimes N(\eta)_1 \rho(\alpha)),\\
~Tr(M^1_0 \otimes N^0_0 \rho(\alpha))= Tr(M^1_0 \otimes N(\eta)_0 \rho(\alpha)),\\
~Tr(M^1_1 \otimes N^0_0 \rho(\alpha))= Tr(M^1_1 \otimes N(\eta)_0 \rho(\alpha)),\\
~Tr(M^1_0 \otimes N^0_1 \rho(\alpha))= Tr(M^1_0 \otimes N(\eta)_1 \rho(\alpha)),\\
~Tr(M^1_1 \otimes N^0_1 \rho(\alpha))= Tr(M^1_1 \otimes N(\eta)_1 \rho(\alpha)).
\end{cases}
\end{equation}
where $M^x$ and $N^a$ are defined as in Eqs.~\eqref{opm}-\eqref{opn} and $\theta_x$ and $\phi_a$ are those in Eq.~\eqref{ops2}. Furthermore, $\rho(\alpha)=| \psi_{\alpha} \rangle \langle \psi_{\alpha} |$ and $|\psi_{\alpha}\rangle=\cos(\alpha) |00 \rangle +\sin(\alpha) |11 \rangle$.
An equivalent way to implement $N(\eta)$ in Eq.~\eqref{n_theta}, which is the one we choose, is to keep the PC in its optical axis and put it in between of two HWPs, rotated of $\frac{\eta}{2}$, so that the input polarization state is rotated of $\eta$ and, after the PC, it is rotated back.
 
\section{Experimental qACE lower bounds}
The curve in Fig.~1a of the main text displays a case where the expected qACE amounts to $0$, for any degree of entanglement $\alpha$, considering a generated state $|\psi\rangle= cos(\alpha) |00\rangle +sin(\alpha) |11\rangle$ and the measurement operators constituting $MS1$. Hence, the lower bound of this quantity is trivially 0, given that, according to its definition, any average causal effect must belong to the interval $(0,1)$. 
From an experimental point of view, this was confirmed by our apparatus, because the lower bounds of the qACE were negative for all of the tested $\alpha$. Therefore, in the experimental figure 5a, we set all of the negative lower bounds to 0. In the table below, we report the trivial experimental values that we experimentally obtained applying Eq.~(10) of the main text. The table also reports the distance, in terms of standard deviations, from $0$, to quantify the confidence, with which we can conclude that those values are lower than $0$. Let us note that the same situation occurred also for the $cACE$ lower bound corresponding to $\alpha=0$, for $MS1$, and for the $qACE$ lower bound corresponding to $\alpha=0.785$, for $MS2$. The experimental values amounted, respectively, to $-0.01040 \pm 0.00010$ and to $-0.1573 \pm 0.0030$.
\begin{center}
\begin{tabular}{ |c|c|c|c|}
\hline
$\alpha$ & $\mathrm{qACE}_{LB}$  & $\pm$&  $ \mathrm{D}_{\sigma}$\\ 
 \hline
 ~~~~~~0~~~~~~   & ~~~-0.35118~~~  & ~~~0.00055~~~  & ~~~642.80~~~\\  
 0.209  & -0.2776  & 0.0027    & 100.72\\
 0.305 & -0.1823 & 0.0046 & 39.72\\
 0.393 & -0.2472 & 0.0024   & 101.90\\
 0.523  & -0.1870 & 0.0036   & 53.49\\
 0.698  & -0.1211 & 0.0042   & 27.95\\
 0.785 & -0.1782 & 0.0046   & 39.42\\
 \hline
\end{tabular}
\end{center}